\documentclass[sigconf]{acmart}

\copyrightyear{2023}
\acmYear{2023}
\setcopyright{acmlicensed}
\setcopyright{rightsretained}
\acmConference[KDD '23] {Proceedings of the 29th ACM SIGKDD Conference on Knowledge Discovery and Data Mining}{August 6--10, 2023}{Long Beach, CA, USA.}
\acmBooktitle{Proceedings of the 29th ACM SIGKDD Conference on Knowledge Discovery and Data Mining (KDD '23), August 6--10, 2023, Long Beach, CA, USA}
\acmPrice{}
\acmISBN{979-8-4007-0103-0/23/08}
\acmDOI{10.1145/3580305.3599902}

\AtBeginDocument{%
  \providecommand\BibTeX{{%
    \normalfont B\kern-0.5em{\scshape i\kern-0.25em b}\kern-0.8em\TeX}}}

\usepackage{tikz}
\usepackage{amsmath}
\usepackage{booktabs}
\usepackage{multirow}
\usepackage{amsmath,amsfonts}
\usepackage[inline,shortlabels]{enumitem}
\usepackage{amsmath}

\newcommand{\ie}{\text{i.e.},~}

\newcommand{\eg}{\text{e.g.},~}

\newcommand{\etal}{\text{et al.}~}

\newcommand{\name}{\textit{AnoFusion}}
\newcommand{\fscore}{$F_{1}$-score}
\newcommand{\point}[1]{\vspace{1mm}\noindent\textbf{#1}.}

\begin{document}

\ccsdesc[500]{Software and its engineering~Maintaining software}
\ccsdesc[500]{Computing methodologies~Failure Detection; Graph Neural Networks}

\title{Robust Multimodal Failure Detection for Microservice Systems}

\settopmatter{authorsperrow=4}

\author{Chenyu Zhao*}
\affiliation{%
  \institution{Nankai University}
  \city{Tianjin}
  \country{China}
  }

\author{Minghua Ma}
\authornote{Equal Contribution}
\affiliation{%
  \institution{Microsoft}
  \city{Beijing}
  \country{China}
}

\author{Zhenyu Zhong}
\author{Shenglin Zhang}
\authornote{Corresponding Author}
\affiliation{%
  \institution{Nankai University}
  \city{Tianjin}
  \country{China}
  }

\author{Zhiyuan Tan}
\author{Xiao Xiong}
\affiliation{%
  \institution{Nankai University}
  \city{Tianjin}
  \country{China}
  }

\author{LuLu Yu}
\author{Jiayi Feng}
\affiliation{%
  \institution{Nankai University}
  \city{Tianjin}
  \country{China}
  }

\author{Yongqian Sun}
\author{Yuzhi Zhang}
\affiliation{%
  \institution{Nankai University}
  \city{Tianjin}
  \country{China}
}

\author{Dan Pei}
\affiliation{%
  \institution{Tsinghua University}
  \city{Beijing}
  \country{China}
}

\author{Qingwei Lin}
\author{Dongmei Zhang}
\affiliation{
  \institution{Microsoft}
  \city{Beijing}
  \country{China}
}

\renewcommand{\authors}{Chenyu Zhao, Minghua Ma, Zhenyu Zhong, Shenglin Zhang, Zhiyuan Tan, Xiao Xiong, LuLu Yu, Jiayi Feng, Yongqian Sun, Yuzhi Zhang, Dan Pei, Qingwei Lin, Dongmei Zhang}
\renewcommand{\shortauthors}{C.Zhao, M.Ma, Z.Zhong, S.Zhang, Z.Tan, X.Xiong, L.Yu, J.Feng, Y.Sun, Y.Zhang, D.Pei, Q.Lin, D.Zhang}

\begin{abstract}
Proactive failure detection of instances is vitally essential to microservice systems because an instance failure can propagate to the whole system and degrade the system's performance.
Over the years, many single-modal (\ie metrics, logs, or traces) data-based anomaly detection methods have been proposed. 
However, they tend to miss a large number of failures and generate numerous false alarms because they ignore the correlation of multimodal data.
In this work, we propose \name{}, an unsupervised failure detection approach, to proactively detect instance failures through multimodal data for microservice systems.
It applies a Graph Transformer Network (GTN) to learn the correlation of the heterogeneous multimodal data and integrates a Graph Attention Network (GAT) with Gated Recurrent Unit (GRU) to address the challenges introduced by dynamically changing multimodal data.
We evaluate the performance of \name{} through two datasets, demonstrating that it achieves the \fscore{} of 0.857 and 0.922, respectively, outperforming the state-of-the-art failure detection approaches.

\end{abstract}

\keywords{Microservice System, Multimodal, Failure Detection}

\maketitle

\section{Introduction}

As an increasing number of Internet applications migrate to the cloud, the microservice architecture, which allows each microservice to be independently developed, deployed, upgraded, and scaled, has attracted widespread attention recently~\cite{2021faultanalysis}.
A microservice system is typically a large-scale system with many \textit{instances} (\eg virtual machines or containers).
Correlations among instances, \eg service invocations and resource contention, are usually complex and dynamic~\cite{yu2021microrank}.
When an instance fails, it may degrade the performance of the whole microservice system, impact user experience and even lead to revenue loss.
For example, some failed instances resulted in a surge of connection activity that overwhelmed the networking devices between the internal network and the main AWS network in December 2021~\cite{aws}.
Therefore, it is crucial to proactively detect instance failures to mitigate failures timely.

Operators continuously collect three types of monitoring data, including metrics, logs, and traces for proactively detecting instance failures~\cite{Multilevel-Observability}.
The metrics include system-level metrics (\eg CPU utilization, memory utilization, and network throughput) and user-perceived metrics (\eg average response time, error rate, and page view count).
A log records the hardware or software runtime information, including state changes, debugging output, and system alerts.
For an API request, a trace records its invocation chain through instances, where each service invocation is called a span.

We adopt failure and anomaly to characterize the faulty behaviors of instances and monitoring data: 1) \textit{an anomaly} is a deviation from the normal system state (often reflected in monitoring data), and 2) \textit{a failure} is an event where the service delivered by an instance goes wrong, and  user experience is degraded~\cite{ma2022empirical}.
Table~\ref{table:sigle modal metric anomaly detection} lists some types of anomalies and failures.
For example, when a ``login failure'' occurs, users cannot log into the system successfully.
Anomalies in logs and traces can be detected when this failure happens: many ``ERROR''s will be printed in logs, and some trace data will have significantly larger Response Time (RT).
It is common to observe many anomalies during a service failure. 
However, an (intermittent) anomaly does not necessarily lead to a failure.

\begin{table}[!ht]
    \centering
    \caption{
        The anomalies of multimodal data during service failures. ``Mem'' represents the memory utilization metric, ``ERR'' is an error log, and $RT_{S_x\rightarrow S_y}$ denotes the response time when service instance $S_x$ calls $S_y$. ``--'' means no anomaly is found or data is lost in that data modality. 
    }
    \label{table:sigle modal metric anomaly detection}
    \begin{tabular}{ccccc}
        \toprule
        \textbf{Failure Type} & \textbf{Metric} & \textbf{Log} & \textbf{Trace} & \textbf{\# Failures} \\
        \midrule
        failed of QR code &  Mem $\uparrow$ & -- &  -- & 505\\
        system stuck & Mem $\downarrow$ & -- & -- & 16 \\
        login failure & -- & ERR & $RT_{S_1\rightarrow S_2}$=11s & 527\\
        file not found & -- & -- & $RT_{S_2\rightarrow S_3}$=1.5s & 36\\
        access denied & -- & ERR & $RT_{S_2\rightarrow S_4}$=1.1s & 15 \\
        \bottomrule
    \end{tabular}
\vspace{-0.2cm}
\end{table}

Over the years, a significant number of methods have been proposed for automatic metric/log/trace (from now on, we call them single-modal) \textit{anomaly detection}.
They try to proactively detect the single-modal data's anomalous behaviors and determine that the instance fails when the monitoring data becomes anomalous.
However, after investigating hundreds of instance failures (see \S~\ref{sec:datasets}), we conclude that previous methods do not work well for \textit{instance failure detection} in microservice systems.
Correlating metrics, logs, and traces (from now on, we call them multimodal) is crucial for instance failure detection.
On the one hand, the single-modal data cannot reveal all types of failures, let alone itself can be missing or collected too slowly \cite{ma2022empirical}.
For example, in Table~\ref{table:sigle modal metric anomaly detection}, when the failure ``failed to generate QR code'' happens, only the instance's metrics, \ie memory utilization, increase dramatically and exhibit anomalous behaviors.
If only conducting log or trace anomaly detection, this type of failure will be falsely ignored.
On the other hand, simply combining the anomaly detection results of the single-modal anomaly detection methods may generate false alarms, which have been confirmed in our experiments (see Table~\ref{table:baselines anomaly detection result}).
For instance, the (transient) anomalies detected by single-modal anomaly detection methods may not represent any instance failure.
Suppose an instance's network throughput metric experiences an anomaly and an alarm is reported because the metric increases suddenly and returns to the normal level after a short period.
However, the system still delivers normal service, because no trace data becomes anomalous, and user experience is not impacted.
Hence, no instance failure should be reported.

To this end, we aim to correlate the multimodal data to detect instance failures for microservice systems, which face the following two challenges.
(1) Modeling the \textit{complex correlations} among multimodal data.
When a failure occurs, one, two, or three modalities of data can become anomalous, and they are correlated with each other. 
Neglecting the correlations can degrade the failure detection accuracy.
(2) Dealing with the \textit{heterogeneous and dynamically changing multimodal data}.
Specifically, metrics are usually in the form of multivariate time series, and logs are typically semi-structured text.
Moreover, a trace consists of spans in a tree structure.
Integrating such heterogeneous multimodal data is quite challenging.
Additionally, an instance's multimodal data usually changes dynamically over time.

In this work, we propose \name{}, an unsupervised instance failure detection approach for microservice systems.
To address the first challenge, we apply  Graph Transformer Network (GTN)~\cite{KipfW17, DBLP:journals/tkde/ZhangCZ22} since it can embed multimodal data into a graph and learn the correlation of heterogeneous data through the effective representations of graph nodes and edges~\cite{DBLP:conf/nips/YunJKKK19,DBLP:conf/aaai/XiaHXDZYPB21}. 
To address the second challenge, we first serialize the data of each modality according to the modality's characteristics and construct the nodes and edges of heterogeneous graphs.
After that, we adopt Graph Attention Network (GAT)~\cite{DBLP:conf/iclr/VelickovicCCRLB18}, which assigns different weights to neighbor nodes and learns the dynamic patterns of multimodal data, to optimize the graphs and filter significant node information.
We then use a Gated Recurrent Unit (GRU)~\cite{ijcai2019-705} to capture the temporal information and predict the multimodal data of the next moment. 
Finally, the similarity between the observation and prediction values is used to determine whether an instance fails.

The contributions of this paper are summarized as follows:
\begin{itemize}[leftmargin=*]
    \item To the best of our knowledge, we are among the first to identify the importance of exploring the correlation of multimodal monitoring data (\ie metrics, logs, and traces), and correlate the multimodal data using GTN for instance failure detection.
    
   \item Our approach, \name{},  serializes the data of the three modalities according to each modality's trait. 
    It combines GTN and GAT to detect anomalies in the dynamic multimodal data robustly. 
    In addition, a GRU layer is used to capture the temporal information of the multimodal data.
    
    \item
    We adopt two microservice systems, consisting of 10 and 28 instances, respectively, to evaluate the performance of \name{}.
    The evaluation results show that \name{}  detects instance failures with average \fscore{} of 0.857 and 0.922, outperforming baseline methods by 0.278 and 0.480 on average, respectively.
\end{itemize}

Our source code and experimental data are available at \textbf{\url{https://github.com/zcyyc/AnoFusion}}.
\section{Background and Related Work}

\subsection{Single-modal Anomaly Detection}
\label{multimodal data of cloud-na}

\begin{figure}
    \begin{center}
    \includegraphics[width=.6\linewidth]{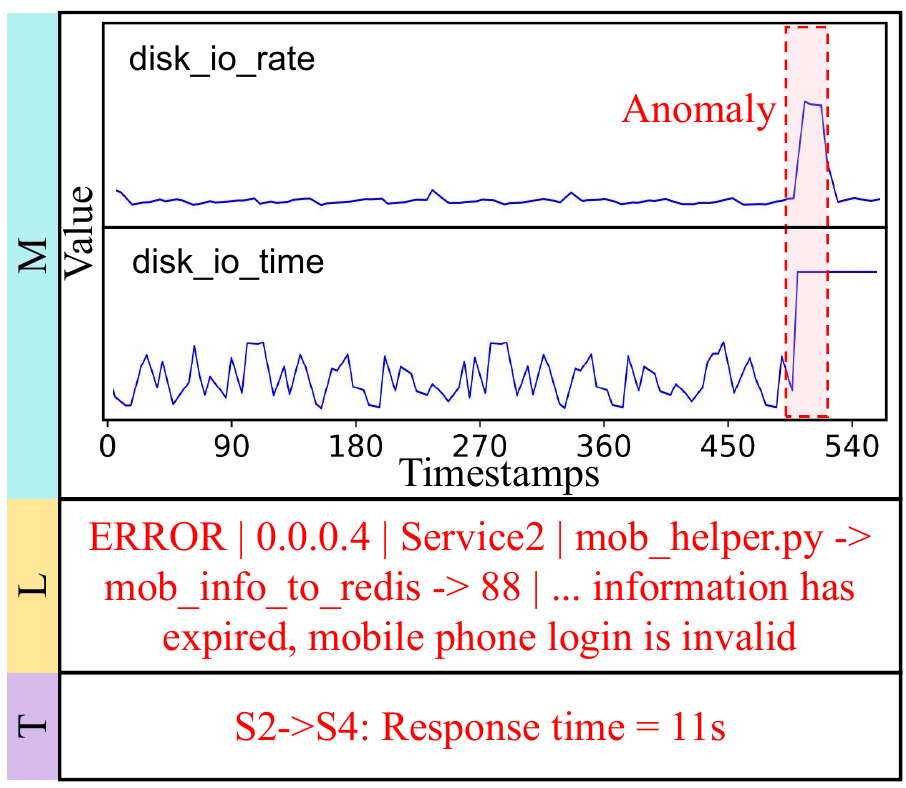}
    \end{center}
    \caption{The multimodal monitoring data, \ie \textbf{M}etrics, \textbf{L}ogs, and \textbf{T}race, during an instance failure}
    \label{fig:correlation and heterogeneity}
\end{figure}

Generally, operators continuously collect three types of observable monitoring data: metrics, logs, and traces~\cite{Multilevel-Observability} to ensure the reliability of microservice systems.
Figure~\ref{fig:correlation and heterogeneity} shows an example of anomalous multimodal data in a failure case.

\point{Metric}
A metric is defined as $\mathbf{x}=\left\{x_{1},x_{2},\ldots,x_{T}\right\}$, where $T$ is the length of the metric, $x_{t}\in\mathbb{R}$ denotes the observation at time $t$. 
A microservice instance typically has a set of metrics that can be represented as a multivariate time series, monitoring various service metrics (\eg page view) and system/hardware metrics (\eg CPU usage).
Figure~\ref{fig:correlation and heterogeneity} (M part) shows an example of metric data.
Traditional statistic metric anomaly detection methods~\cite{ma2020diagnosing} do not need training data but can be less effective when facing intricate data.
Supervised learning methods~\cite{liu2015opprentice, laptev2015generic} need operators to manually label anomalies, which is impractical in many real-world scenarios.
Thus, unsupervised methods ~\cite{DBLP:conf/kdd/AudibertMGMZ20, DBLP:conf/usenix/MaZ0XLLNZWP21, DBLP:conf/usenix/MaZ0XLLNZWP21,
zhang2019robust,  
su2019robust} that do not require anomaly labels have become a hot research topic in recent years.
For example, 
 JumpStarter~\cite{DBLP:conf/usenix/MaZ0XLLNZWP21} applies a compressed sensing technique for anomaly detection. 
USAD~\cite{DBLP:conf/kdd/AudibertMGMZ20} detects anomalies through adversarial training with high efficiency.
A metric anomaly detection method can easily detect an instance failure if multiple metrics become anomalous soon after the failure. 
However, since metric anomaly detection methods only utilize metric data to detect anomalies and possible failures in a system, they will fail to alert operators when a failure does not manifest itself on metrics.
\name{} analyzes metric data in an unsupervised way and reduces false alarms by using metric data together with other data modalities.


\point{Log}
Log data is semi-structured text output by instances at the application or system level.
It is typically used to record the operational status of hardware or software.
Generally, logs are generated with a predefined structure.
As a result, extracting log templates and their parameters is a standard step in analyzing log data~\cite{fu2022investigating}.
For example, Figure~\ref{fig:correlation and heterogeneity} (L part) lists a log.
Traditional log anomaly detection methods are usually designed to identify keywords in logs like ``ERROR'' or ``fail''.
However, negative keywords such as ``fail'' may appear in logs due to network jitters or operator login failure, and they do not imply an instance failure.
Advanced approaches follow a similar workflow: log parsing, feature extraction, and anomaly detection \cite{he2022empirical}.
Deep learning-based methods learn the log patterns (\eg sequential feature, quantitative relationship) of normal executions and determine an anomaly when the pattern of a log sequence deviates from the learned normal patterns~\cite{ying2021improved, du2017deeplog, liu2022uniparser}.
For example, LogAnomaly~\cite{meng2019loganomaly} applies template vectors to extract the hidden semantic information in the log templates and detects continuous and quantitative log anomalies  at the same time.
    Deeplog~\cite{du2017deeplog} predicts the logs that may appear after a sliding window utilizing the LSTM model.
\name{} requires neither labeling work nor domain knowledge when analyzing log data.

\point{Trace}
A trace is made up of spans, each of which corresponds to a service invocation~\cite{nedelkoski2019anomaly}.
Figure~\ref{fig:correlation and heterogeneity} (T part) shows an example of trace data.
When the service processes a user's request, several instances will be invoked.
The monitoring system records when a specific service is called and when it responds, and the difference between them is the Response Time (RT).
Most trace anomaly detection methods detect anomalies according to whether the response time of each invocation increases dramatically and/or whether the invocation path behaves abnormally~\cite{liu2020unsupervised, lipractical, nedelkoski2019anomaly, 2020graph-based,nedelkoski2019anomaly1}.
For instance, TraceAnomaly~\cite{liu2020unsupervised}  learns the normal patterns of traces, and anomalies are detected when their patterns deviate from those of normal traces.
However, on the one hand, a trace anomaly alone does not necessarily denotes an instance fails.
On the other hand, an instance failure may not manifest itself in the trace data.
Therefore, using trace anomaly detection methods alone can also lead to missed alerts or false alarms.
\name{} can combine trace data with other modalities to boost anomaly detection performance.

\subsection{Multimodal Anomaly Detection}

Deep learning-based multimodal data fusion has witnessed great success in several research fields.
For example, video subtitle generation~\cite{iashin2020multi}, conversation behavior classification~\cite{saha2020towards}, and emotion recognition~\cite{DBLP:conf/mm/JiaLWFXC21}. 
Recent studies have started to tackle the anomaly detection problem based on multimodal data.
Vijayanand \etal\cite{DBLP:journals/jifs/VijayanandS22} propose an anomaly detection framework for cloud servers using multidimensional data, including different features such as network traffic sequence features, CPU usage, and memory usage from host logs. 
These extracted multidimensional features are fed to the detection model that identifies the anomalies and maximizes the detection accuracy.
\cite{DBLP:journals/jss/FarshchiSWG18} performs correlation analysis on metrics and logs to discover the anomaly patterns in cloud operations.
Additionally, SCWarn~\cite{zhao2021identifying} combines metrics and logs for anomaly detection by serializing the metrics and logs separately and adopting LSTM to detect failures.
However, traces, which are vital to instances, are missing in these works, and thus, they cannot achieve optimal performance when detecting anomalies in our scenario.
To the best of our knowledge, we are among the first to focus on detecting instance failures using multimodal data.


\subsection{Graph Neural Networks}
\label{sec:preliminary}

\point{GTN}
GTN takes a heterogeneous graph as input and turns it into a new graph structure specified by meta-paths.
Meta-paths are relational sequences that connect pairs of objects in heterogeneous graphs, which are commonly employed to extract structural information.
By combining multiple GT layers with GCN, GTN learns node representations on the graph efficiently in an end-to-end way~\cite{DBLP:conf/nips/YunJKKK19}.
We apply GTN to learn the correlations among multimodal data in our scenario.

\point{GAT} 
GCN is not good at analyzing dynamic graphs, and when the graph structure of training and test sets changes, GCN will no longer be suitable.
In addition, GCN assigns the same weight to each neighbor node, which falls short of our expectations for future graph structure optimization.
GAT solves the problems of GCN by allocating various weights to different nodes.
It enables various nodes to be distinguished in terms of importance, so that \name{} can focus on more significant information in the graph structure~\cite{DBLP:conf/iclr/VelickovicCCRLB18}.
Therefore, GAT is expected to achieve better performance in processing dynamically changing time series data, and thus we utilize GAT instead of GCN.

\point{GRU}
As we know, RNN~\cite{DBLP:conf/aaai/KimACP21} can represent time dependency by adopting deterministic hidden variables.
However, RNN may be incapable of dealing with the long-term dependency problem in the time series, and LSTM~\cite{DBLP:conf/acl/JiaZ20} and GRU~\cite{ijcai2019-705} are proposed as solutions.
Generally, GRU is often comparable to LSTM, and the fewer parameters and more straightforward structure make it ideal for model training~\cite{su2019robust}. We thus apply GRU to capture the time dependency of the multimodal data.

\section{Motivation}
\label{empirical}

To prove that single-modal data is insufficient to comprehensively capture the failure patterns of instances, we adopt two datasets (see Section~\ref{subsec:experimental setup} for more details) for an empirical study.
These datasets contain the multimodal data (\ie metrics, logs, and traces) collected from microservice systems.
It also includes the records of all failure injections for a fair evaluation.

We perform a thorough internal data analysis to investigate the correlation of different modalities.
Table~\ref{table:sigle modal metric anomaly detection} lists some monitoring data collected from a microservice system.
Many instance failures cannot be successfully captured using single-modal data.
It also shows that when a failure occurs, data of different modalities may display anomalous patterns at the same time.
Mining the correlation between multimodal data can provide more comprehensive and accurate information for failure detection tasks.

Moreover, we experiment to evaluate the failure detection performance of single-modal data-based anomaly detection methods.
We apply five popular single-modal anomaly detection methods (Section~\ref{subsec:experimental setup}),
JumpStarter~\cite{DBLP:conf/usenix/MaZ0XLLNZWP21}, USAD~\cite{DBLP:conf/kdd/AudibertMGMZ20}, LogAnomaly~\cite{meng2019loganomaly}, Deeplog~\cite{du2017deeplog}, TraceAnomaly~\cite{liu2020unsupervised}, and the combination of JumpStarter, LogAnomaly, and TraceAnomaly, to conduct metric/log/trace anomaly detection, respectively.
Table~\ref{table:baselines anomaly detection result} lists the precision, recall, and \fscore{} of these methods.

\point{Metric anomaly detection}  JumpStarter and USAD achieve low performance on the two datasets.
JumpStarter extracts real-time normal patterns from metric data, but it does not consider the patterns of historical data.
USAD is not very noise-resilient, which results in a significant number of false positives and false negatives.
Furthermore, they do not take logs and traces into account, thus they lack essential information from logs and traces for failure detection tasks.

\point{Log anomaly detection} LogAnomaly and Deeplog achieve relatively high \fscore{} on D1.
It is because the anomaly patterns of the log data in D1 are more obvious and straightforward to identify than those in D2.
When a keyword like ``ERROR'' appears in logs, there is a high probability that it is an instance failure.
However, only relying on log data results in a large number of false positives and false negatives in D2, for some failures do not manifest themselves obviously in logs, and some anomalous logs do not indicate an instance failure.

\point{Trace anomaly detection} TraceAnomaly gets unsatisfactory \fscore{} on both datasets.
The precision of TraceAnomaly is low, indicating that there are a huge number of false positives.
Because TraceAnomaly determines anomalies based on only response time.
However, a larger response time quickly returning to normal status does not indicate an instance  failure.

\point{JLT} JLT aggregates the results of \textbf{J}umpStarter, \textbf{L}ogAnomaly, and \textbf{T}raceAnomaly directly, to form a multimodal baseline.
    The aggregation method is majority voting, which determines a failure if two or more modalities have anomalies at a certain moment.
    JLT suffers from low precision (recall) on D1 (D2), indicating a high false positive (negative) rate,  because it ignores the correlation of the multimodal data.
    On the one hand, some failures only manifest themselves in a specific modality of data, making the majority voting strategy requiring two or more modalities to have anomalies for failure detection ineffective, which results in many false negatives on D2.
    On the other hand, since JumpStarter, LogAnomaly, and TraceAnomaly all suffer from low precision on D1, their combination, \ie JLT, still experiences a high false positive rate.
    

In summary, single-modal anomaly detection approaches fail to detect failures robustly since they lack insights from other data modalities.
Additionally, simply combining the anomaly detection results of the single-modal anomaly detection methods, instead of mining the correlations of the multimodal data, cannot guarantee high accuracy.
Therefore, we attempt to design a robust instance failure detection approach by correlating the multimodal data.

\begin{figure*}[t]
    \centering
    \includegraphics[width=\linewidth]{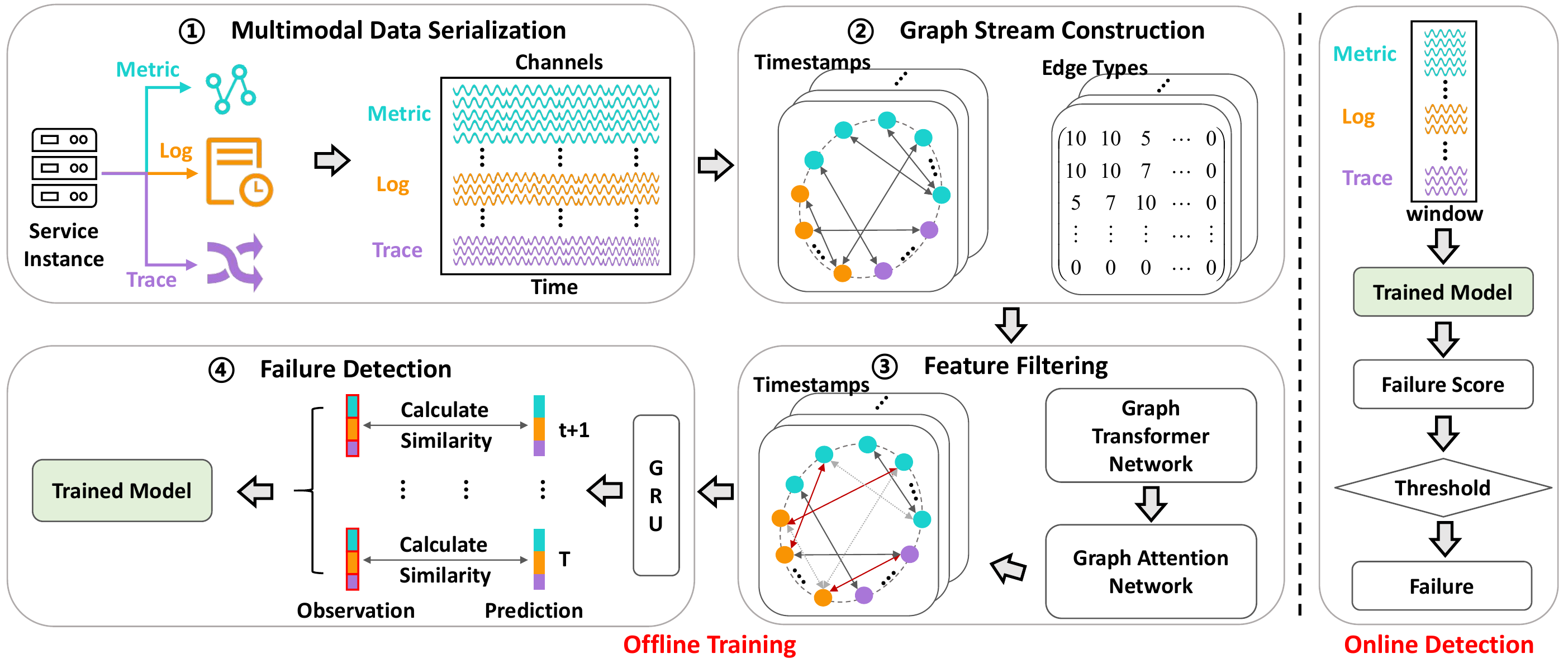}
    \caption{
        The framework of \name{}. It is an unsupervised learning approach without using labels in the offline training. 
    }
    \label{fig:framwork}
\end{figure*}

\section{\name{}}

\subsection{Overview}
\label{overview}

As shown in Figure~\ref{fig:framwork}, the workflow of \name{} is divided into the offline training stage and the online detection stage.
To capture
the heterogeneity and correlation among multimodal data, we employ GTN to update the heterogeneous graphs.
Moreover, to improve the robustness of \name{}, we apply GAT after GTN, making it perform stably when the data patterns of the training set and test set are different.
In addition, to achieve unsupervised failure detection and make \name{} more suitable for time series data, we use GRU to predict the multimodal data of the next moment.
In the offline training stage, \name{} consists of four main steps:

\begin{itemize}[leftmargin=*]
    \item \textbf{Multimodal Data Serialization.}
    To prepare for the future graph structure's construction, \name{} converts multimodal data (i.e., metrics, logs, and traces) into time series using predefined processes and aligns their time.
    After serialization, \name{} treats each time series as a ``data channel''.
    \item \textbf{Graph Stream Construction.}
    To build the raw inputs for GTN, \name{}  constructs a heterogeneous graph containing all the data channels based on their connections for each moment $t$.
    Then, all heterogeneous graphs form a graph stream, which will be input into GTN.
    \item \textbf{Feature Filtering.}
    GTN updates the graph stream by learning the representations of nodes in the heterogeneous graph and capturing the correlation among different data modalities.
    The updated graph stream is regarded as the features of the original data channels.
    Then, \name{} utilizes GAT to give attention scores to the nodes in the graph stream, identifying different patterns and achieving feature filtering.
    \item \textbf{Failure Detection.}
    GRU is applied to temporal sequences to predict the values at the next moment based on the previous inputs.
    We train the GRU network to predict the next graph based on the given graph stream as accurately as possible.
\end{itemize}

In the online detecting stage, for a given time $t$, multimodal data will be serialized according to the observations in $[t-\theta,t]$, where $\theta$ is the input window size.
Then, we use the serialized data channels to construct a heterogeneous graph stream.
The graph stream of this window will be fed into the trained model to obtain the prediction of the next graph.
\name{} calculates the similarity between the predicted graph and the observed graph as the failure score and then determines whether it is a failure.
Note that \name{} does not restrict the dimensions of any modality data.

\subsection{Multimodal Data Serialization}
\label{serialize}

\textbf{Serialization of metric data.}
Metrics collected from instances are in the form of time series, which have a serialized structure.
Therefore, it only requires regular preprocessing steps such as normalization.
The normalization process is given by:
    $ \hat{\mathbf{m}}\equiv{\mathbf{m}} / {\lvert\mathbf{m}\rvert} $
where $\mathbf{m}$ is the raw metric data and $\hat{\mathbf{m}}$ is the normalized data.
It scales individual samples to have a unit norm, which can be useful when using a quadratic form such as the dot-product to quantify the similarity of any pair of samples.


\point{Serialization of log data}
Parsing logs correctly and extracting log templates are the two essential steps of log serialization~\cite{wang2020RootCause}.
We adopt the advanced log parsing algorithm, Drain~\cite{he2017drain}, which has shown its superiority and efficiency, to extract log templates in \name{}.
The log serialization process is the following two steps:

\noindent (1) \textbf{Clustering.} 
To deal with log changes caused by constantly updating code, adding new logs, and altering new logs in actual microservice systems, we first use a clustering algorithm for the log templates.
By grouping similar log templates into clusters, on the one hand, the redundant information can be removed, and on the other hand, the types of log templates can be used to characterize log data. 
Once a new log template emerges due to a software update, the similarity between the new log template and the previous cluster centroids can be calculated, and it can be decided whether the new log template belongs in an existing cluster or should be regarded as a new cluster.
Furthermore, through the empirical study of a large number of online service systems, we conclude that failures rarely occur in real-world scenarios~\cite{DBLP:conf/usenix/MaZ0XLLNZWP21}.
Since~\name{} is an unsupervised learning method based on the assumption that all training samples have normal patterns, removing anomalous log templates will improve the model's performance.
Based on the analysis mentioned above, we finally utilize the ``bert-base-uncased'' model
~\cite{DBLP:conf/naacl/DevlinCLT19} to obtain sentence embedding vectors, and apply the DBSCAN~\cite{DBSCAN} algorithm to cluster these vectors.
\name{} computes the centroid $\mathbf{c}$ of each cluster $\mathcal{C}$ by:
\begin{equation}
    \mathbf{c}=\arg\min_{\mathbf{a}\in\mathcal{C}}\sum_{\mathbf{b}\in\mathcal{C}}\lvert\mathbf{a}-\mathbf{b}\rvert
\end{equation}

\noindent (2) \textbf{Serialization.} The category of each log entry in the input can be determined by calculating the distance between its sentence embedding vector and that of the centroid of each cluster.
After that, \name{} uses a sliding window to split the input log data into windows, each of which has the window length $\theta$ and the step size $\delta$.
We count the number of each category of logs as well as the total number of logs in each window to form $M+1$ time series, where $M$ is the number of log template clusters.
The horizontal axis $Timestamp$ corresponds to the input log's collection time.


\point{Serialization of tracing data}
\name{} uses the sliding window with length $\theta$ and step size $\delta$ to split tracing data.
Each window contains tracing data (in the form of RT) of the invocations related to the instance.
Then, for each window, \name{} computes the mean, median, range, and standard deviation of the invocations RT, producing four time series, respectively.
If status code is available, \name{} may take them as the fifth time series.
We treat each time series as a data channel, similar to the serialization of log data.

\point{Clock synchronization}
To build the graph structure more conveniently, \name{} synchronizes the clocks of the three modal data after serialization.
The goal of \name{} is failure detection for a single instance and all the monitoring data acquired is within that instance.
Therefore, the monitoring data clocks of the three modalities are relatively synchronized.
The metric data is collected every minute. A log entry is generated when an event occurs in the
instance. Moreover, a trace is recorded when a request is processed. Therefore, we obtained
the features (e.g. the number of occurrences) of metrics, logs, and traces every minute in our scenario.

\subsection{Graph Stream Construction}
\label{build graph}

The data channels we get from the previous step can be described as $X=\left\{\mathbf{x}^{\left(1\right)},\ldots,\mathbf{x}^{\left(N\right)}\right\}$, where $N$ is the number of data channels.
\name{} constructs a heterogeneous graph $G_{t}$ for each timestamp $t$ using the extracted data channels.
The node set of graph $G_{t}$, denoted by $X_{t}$, consists of the value of each data channel at time $t$, i.e., $X_{t}=\left\{x_{t}^{\left(1\right)},\ldots,x_{t}^{\left(N\right)}\right\}$.
Since there are three modalities, there are also three types of nodes (i.e., metrics, logs, and traces).
Thus, the number of edge types $K=6$ (i.e., metrics-metrics, metrics-logs, metrics-traces, logs-logs, log-traces, traces-traces).
The adjacency matrix for each type of edge in graph $G_{t}$ can now be expressed as $A^{\left(k\right)}\in\mathbb{R}^{N\times N}$, where $k=1,\ldots,K$.

\name{} utilizes the mutual information (MI)~\cite{DBLP:conf/nips/GaoKOV17} to calculate the adjacency matrix.
For each data channel pair $\left(\mathbf{x}^{\left(i\right)},\mathbf{x}^{\left(j\right)}\right)$ with an edge type of $k$, the corresponding adjacency matrix value can be calculated as follows:
\begin{equation}
    A^{\left(k\right)}_{i,j}=A^{\left(k\right)}_{j,i}=\sum_{a=1}^{\tau}\sum_{b=1}^{\tau}p\left(x_{a}^{\left(i\right)},x_{b}^{\left(j\right)}\right)log\frac{p\left(x_{a}^{\left(i\right)},x_{b}^{\left(j\right)}\right)}{p\left(x_{a}^{\left(i\right)}\right)p\left(x_{b}^{\left(j\right)}\right)}
\end{equation}
where $\tau$ is the number of timestamps (i.e., the length of each data channel), $p\left(\mathbf{x}^{\left(i\right)},\mathbf{x}^{\left(j\right)}\right)$ is the joint probability mass function of $\mathbf{x}^{\left(i\right)}$ and $\mathbf{x}^{\left(j\right)}$, and $p\left(\mathbf{x}^{\left(i\right)}\right)$ and $p\left(\mathbf{x}^{\left(j\right)}\right)$ are the marginal probability mass functions of $\mathbf{x}^{\left(i\right)}$ and $\mathbf{x}^{\left(j\right)}$, respectively.
After calculating MI for all channel pairs, we now have the final adjacency matrix $A\in\mathbb{R}^{N\times N\times K}$.
$G_{t}=\left(X_{t},A\right)$ is defined to be the heterogeneous graph generated at time $t$.
\name{} stacks the graphs of each moment together to form a graph stream $G=\left\{G_{1},\ldots,G_{\tau}\right\}$.

\subsection{Feature Filtering}
\label{feature filtering}

\name{} performs feature filtering by updating the heterogeneous graph stream $G$ with GTN and learning failure patterns by GAT.

\point{Graph Transformer Network}
GTN models the heterogeneity and correlation of multimodal channels using the adjacency matrix $A$.
Graph Transformer layers (GT layers) are the main component of GTN.
They learn the soft selection and composite relationship of edge types to produce useful multi-hop connections, also known as meta-path~\cite{DBLP:conf/nips/YunJKKK19}.
Specifically, considering the adjacency matrix $A$ as the input, a GT layer has two steps:
First, it softly constructs several graph structures from $A$ by a $1\times1$ convolutional layer, which can be formulated as:
\begin{equation}
    Q^{\left(k\right)}=\phi\left(A,W^{\left(k\right)}\right)=\sum_{i=1}^{K}w_{i}^{\left(k\right)}A^{\left(i\right)}
\end{equation}
where $Q^{\left(k\right)}$ is the generated graph for edge type $k$, $\phi$ denotes the $1\times1$ convolution, $W^{\left(k\right)}\in\mathbb{R}^{1\times1\times K}$ is the parameter of $\phi$ (for edge type $k$), $K$ is the number of edge types.
Second, it combines each $Q^{\left(k\right)}$ through matrix multiplication to generate a new graph structure $A^{\prime}$, a.k.a, meta-path:
\begin{equation}
    A^{\prime}=D^{-1}\prod_{k=1}^{K}Q^{\left(k\right)}
\end{equation}
Note that we also normalize $A^{\prime}$ by $D$, which denotes the degree matrix of $A$, to ensure numerical stability.
Stacking several GT layers in GTN aims to learn a high-level meta-path that is a useful relationship of multimodal data.

\point{Graph Attention Network}
With the meta-path matrix $A^{\prime}$ generated by stacking multiple GT layers, \name{} employs GAT on the heterogeneous graph stream to distinguish the significance of multimodal data channels and completes the feature filtering.
The multi-head attention mechanism is utilized as well to stabilize this process.
Specifically, for each channel pair $\left(\mathbf{x}^{\left(i\right)},\mathbf{x}^{\left(j\right)}\right)$, we first compute a raw attention score for each attention head based on $A^{\prime}$:
\begin{equation}
    \beta_{i,j}^{\left(h\right)}=\mathop{LeakyRelu}\left(A_{i,j}^{\prime}\cdot\mathop{concat}\left(W\mathbf{x}^{\left(i\right)},W\mathbf{x}^{\left(j\right)}\right)\right)
\end{equation}
where $\beta^{\left(h\right)}$ is the attention score for the $h$-th attention head, $W$ denotes the learnable parameter of a linear transformation.
Then, \name{} normalizes the raw attention score with softmax and performs node feature aggregation for the $i$-th node $x^{\left(i\right)}$ by:
\begin{equation}
    \begin{aligned}
         & \widetilde{\beta}_{i,j}^{\left(h\right)\prime}=\mathop{softmax}\left(\beta_{i,j}^{\left(h\right)\prime}\right)=\frac{\exp\left(\beta_{i,j}^{\left(h\right)}\right)}{\sum_{l=1}^{N}\exp\left(\beta_{i,l}^{\left(h\right)}\right)} \\
         & \mathbf{x}^{\left(i\right)\prime}=W_{H}\cdot\mathop{concat}\limits_{h=1}^{H}\left(\sum_{j=1}^{N}W^{\left(h\right)}\mathbf{x}^{\left(j\right)}\widetilde{\beta}_{i,j}^{\left(h\right)\prime}\right)
    \end{aligned}
\end{equation}
where $H$ is the number of attention heads, $\beta_{i,l}^{(h)}$ denote the $h$-th head attention score between channel $i$ and channel $l$, $W^{\left(h\right)}$ and $W_{H}$ denote the linear transformation for each head and final output, respectively.
The data channels are successively updated across the multi-layer Graph Attention Network.

\subsection{Failure Detection}
\label{anomaly detection}
After feature filtering, we use the updated graph stream to train a failure detection model based on a recurrent neural network.
Let $X^{\prime}\in\mathbb{R}^{N\times \tau}$ denote the updated data channels, we can use GRU to capture its complex temporal dependence and predict the value of data channels at time $\tau$.
The GRU network can be formulated as:
\begin{equation}
    \begin{aligned}
        z_{t}       & =\sigma\left(W_{z}X_{t}^{\prime}+U_{z}h_{t-1}+b_{z}\right)                        \\
        r_{t}       & =\sigma\left(W_{r}X_{t}^{\prime}+U_{r}h_{t-1}+b_{r}\right)                        \\
        \hat{h}_{t} & =\tanh\left(W_{h}X_{t}^{\prime}+U_{h}\left(h_{t-1}\odot r_{t}\right)+b_{h}\right) \\
        h_{t}       & =\left(1-z_{t}\right)\odot h_{t-1}+z_{t}\odot\hat{h}_{t},
    \end{aligned}
\end{equation}
where $\sigma$ denotes the sigmoid function, $\odot$ denotes the Hadamard product (i.e., element-wise product), $X_{t}^{\prime}$ is the input vector, $h_{t-1}$ is the previous hidden state, $\hat{h}_{t}$ is the candidate activation vector, $h_{t}$ is the hidden state and output vector of time $t$.
$z_{t}$ denotes the update gate, controlling how much information $h_{t}$ needs to keep from $h_{t-1}$, and how much information needs to be received from $\hat{h}_{t}$.
$r_{t}$ denotes the reset gate, controlling whether the calculation of the candidate activation vector depends on the previous hidden state.
$W$ and $U$ are trainable parameter matrices, and $b$ is a trainable parameter vector.
\name{} uses the final hidden state of GRU, $h_{t}$, to predict the value of data channels at time $\tau$ (i.e., the last moment in the graph stream):
\begin{equation}
    \hat{X}_{\tau}^{\prime}=\tanh\left(W_{o}h_{\tau-1}+b_{o}\right)
\end{equation}
where $W_{o}$ and $b_{o}$ are the learnable parameters.
\name{} adopts mean squared error (MSE) between the predicted value $\hat{X}_{\tau}^{\prime}$ and the observation $X_{\tau}^{\prime}$ as the loss function:
\begin{equation}
    \mathcal{L}=\frac{1}{N}\Vert\hat{X}_{\tau}^{\prime}-X_{\tau}^{\prime}\Vert_{2}^{2}
\end{equation}
where $N$ is the number of data channels.
The GRU network is updated using this loss function iteratively.

\subsection{Online Detection}
\label{online}

In the online detection stage, for a new-coming multimodal monitoring data $X_{t}$, \name{} will first serialize the data using its previous historical observations, i.e., $X_{t-\theta+1:t-1}$, and construct the graph stream $G=\left\{G_{t-\theta+1},\ldots,G_{t}\right\}$, where $\theta$ is the length of the window.
Then, the graph stream is fed into the trained model to get a prediction $\hat{X}_{t}$ for $X_{t}$.
We calculate the difference between the observed and predicted values for each data channel $n$~\cite{deng2021graph}:
\begin{equation}
    ERR_{n}=\lvert\hat{X}_{t}^{\left(n\right)}-X_{t}^{\left(n\right)}\rvert
\end{equation}
Failures may only happen in part of the multimodal data, so we focus on the biggest error.
\name{} utilizes the max function to aggregate $ERR_{n}\left(t\right),n\in\left[1,N\right]$:
\begin{equation}
    ERR=\max_{n=1}^{N}\frac{ERR_{n}-\widetilde{\mu}}{\widetilde{\sigma}}
\end{equation}
where $ERR$ is the failure score at time $t$, $\widetilde{\mu}$ and $\widetilde{\sigma}$ are the median and inter-quartile range (IQR) of the set composed by $ERR_{n}$, respectively.
We use median and IQR instead of mean and standard deviation as they are more robust.

\name{} uses a threshold to determine if a failure has occurred at a specific time $t$.
However, using a static threshold is not effective since data distribution changes over time.
To solve this problem, we employ the Extreme Value Theory (EVT)~\cite{siffer2017anomaly} to automatically and accurately determine the threshold.
EVT is a statistical theory that identifies the occurrences of extreme values and doesn't make any assumptions about data distribution.
EVT can be applied to estimate the likelihood of observing the extreme value for anomaly detection.
EVT has been shown to accurately choose thresholds in previous failure detection methods~\cite{ma2018robust, DBLP:conf/usenix/MaZ0XLLNZWP21}.

\section{Evaluation}
\label{evaluation}
In this section, we address the following research questions:
\begin{itemize}[leftmargin=*]
\item \textbf{RQ1:} How well does \name{} perform in failure detection?
\item \textbf{RQ2:} Does each component contribute to \name{}?
\item \textbf{RQ3:} How do the major hyperparameters of \name{} influence
its performance?
\end{itemize}

\subsection{Experimental Design}
\label{subsec:experimental setup}

\subsubsection{Datasets}
\label{sec:datasets}

To evaluate the performance of \name{}, we conduct extensive experiments using two microservice systems (forming dataset 1 and 2, respectively).
Table~\ref{table:first dataset} lists the detailed information of the datasets.
The second column indicates the number of microservices of each dataset.
The third column indicates the number of instances of each dataset.
The fourth column displays the average failure ratio ($\frac{The~number~of~failure~windows}{The~total~number~of~windows}$) of all instances. 
The fifth column lists every modality, and the last column shows the number of metrics, logs, or traces.

\begin{table}
    \centering
    \caption{
       The detailed information of the two datasets.
       \#Micro and \#Ins denote the number of microservices and instances, respectively.
    }
    \label{table:first dataset}
    \small
    \begin{tabular}{c|c|c|c|c|c}
        \toprule
        & \textbf{\#Micro} & \textbf{\#Ins} & \textbf{\%Failures} & \textbf{Modality} & \textbf{\#} \\
        \midrule
        \multirow{3}*{D1} & \multirow{3}*{5} & \multirow{3}*{10} & \multirow{3}*{4.908} & Metric &  734,165\\
        & & & & Log  &  87,974,577\\
        & & & & Trace &  28,681,438\\
        \midrule
        \multirow{3}*{D2} & \multirow{3}*{14} & \multirow{3}*{28} & \multirow{3}*{1.243} & Metric & 3,122,168\\
        & & & & Log  &  14,894,069\\
        & & & & Trace &  9,473,763\\ 
        \bottomrule
    \end{tabular}
    \vspace{-3mm}
\end{table}

\begin{itemize}[leftmargin=*]
\item \textbf{Dataset 1} (D1) is Generic AIOps Atlas (GAIA) dataset from CloudWise\footnote{\url{https://github.com/CloudWise-OpenSource}}. 
It contains the multimodal data collected from a system
consisting of 10 instances, which is collected more than 0.7 million metrics, 87 million logs, and 28 million traces in two weeks.
Real-world failures are injected, and Table~\ref{table:sigle modal metric anomaly detection} lists some typical symptoms of failure types,
such as QR code generation failure, system stuck, file not found, and access denied, \textit{etc}.

\item \textbf{Dataset 2} (D2) is collected from a large-scale microservice system operated by a commercial bank.
The system has 28 instances such as Web servers, application servers, databases, etc., and provides services for millions of users daily.
Failures are injected into the system manually by professional operators and the multimodal monitoring data (i.e. metrics, logs, and traces) is collected.
In general, these failures can be resource (CPU, memory, disk) failures, network failures (network packet loss and network latency), and application failures (VM failures).
Due to the non-disclosure agreement, we cannot make this dataset publicly available.

\end{itemize}





\subsubsection{Implementation}
\name{} is implemented in PyTorch and all of the experiments are conducted on a Linux Server with two 16C32T Intel(R) Xeon(R) Gold 5218 CPU @ 2.30 GHz, two NVIDIA(R) Tesla(R) V100S, and 192 GB RAM.
In the multimodal data serialization stage, we set the sliding window length $\theta=60$ and step size $\delta=1$ (more discussions can be found in Section~\ref{subsec:eva hyper-parameters}).
In the graph stream construction stage, we set the number of GT layers in GTN to 5, as suggested by~\cite{DBLP:conf/mm/JiaLWFXC21}.
For GAT, the total number of attention heads $H$ is 6 (see Section~\ref{subsec:eva hyper-parameters} for more details).
We split the multimodal monitoring datasets into a training set and a testing set, where the training set contains the front 60\% data of each instance and the testing set contains the rest 40\%.

\subsubsection{Baseline Approaches}
\label{sec:baseline approaches}
We utilize {JumpStarter}~\cite{DBLP:conf/usenix/MaZ0XLLNZWP21}, {USAD}~\cite{DBLP:conf/kdd/AudibertMGMZ20}, {LogAnomaly}~\cite{meng2019loganomaly}, 
 {Deeplog}~\cite{du2017deeplog}, {TraceAnomaly}~\cite{liu2020unsupervised}, {SCWarn}~\cite{zhao2021identifying}, and JLT (see Section~\ref{empirical}), which use single modality, two modalities, or three modalities of data for instance failure detection, as baselines. 
For all approaches, we use grid-search to set their parameters best for accuracy.

\begin{table*}
    \centering
    \caption{
        The average precision, recall, and \fscore{} of different approaches on the two datasets 
    }
    \label{table:baselines anomaly detection result}
    \begin{tabular}{c|ccc|ccc|ccc}
        \toprule
        \multirow{2}{*}{\textbf{Approach}} & \multicolumn{3}{c|}{\textbf{Modality}} & \multicolumn{3}{c|}{\textbf{D1}} & \multicolumn{3}{c}{\textbf{D2}} \\
        \cmidrule{2-10}
        & \textbf{Metric} & \textbf{Log} & \textbf{Trace} & \textbf{Precision} & \textbf{Recall} & \textbf{\fscore{}} & \textbf{Precision} & \textbf{Recall}  & \textbf{\fscore{}} \\
        \midrule
        JumpStarter~\cite{DBLP:conf/usenix/MaZ0XLLNZWP21} & \checkmark &  &   &0.466 & 0.785 & 0.584 & 0.533 & 0.413 & 0.465 \\
        USAD~\cite{DBLP:conf/kdd/AudibertMGMZ20} & \checkmark &  &   & 0.459 & 0.825 & 0.590 & 0.837 & 0.341 & 0.484 \\
        LogAnomaly~\cite{meng2019loganomaly} &  & \checkmark &  & 0.486 & 0.957 & 0.644 & 0.126 & 0.344 & 0.184 \\
        Deeplog~\cite{du2017deeplog} &  & \checkmark &  & 0.506 & 0.812 & 0.623 & 0.105 & 0.275 & 0.151 \\
        TraceAnomaly~\cite{liu2020unsupervised} &  &  & \checkmark & 0.550 & 0.548 & 0.549 & 0.521 & 0.699 & 0.597 \\
        \midrule
        SCWarn~\cite{zhao2021identifying} & \checkmark & \checkmark &  & 0.547 & 0.425 & 0.447 & 0.633 & 0.891 & 0.734 \\
        JLT & \checkmark & \checkmark & \checkmark & 0.461 &  0.940 & 0.618 & 0.800 & 0.344 & 0.481 \\
        \name{} & \checkmark & \checkmark & \checkmark & \textbf{0.795} & \textbf{0.945} & \textbf{0.857} & \textbf{0.863} & \textbf{0.991} & \textbf{0.922} \\
        \bottomrule
    \end{tabular}
\end{table*}

\subsubsection{Evaluation Metrics}
We adopt the widely-used True Positive (TP), False Positive (FP), and False Negative (FN), to label the failure detection results according to the ground truth~\cite{DBLP:conf/usenix/MaZ0XLLNZWP21, ren2019time, ma2018robust}.
A TP is a failure both confirmed by operators and detected by an approach. 
An FP is a normal window that is falsely identified as a failure by an approach.
An FN is a missed failure that should have been detected.
We calculate $precision=TP/\left(TP+FP\right)$, $recall=TP/\left(TP+FN\right)$, 
and $F_{1}$-score$=2\cdot precision\cdot recall/\left(precision+recall\right)$ to evaluate the overall performance of each approach.



\subsection{RQ1: Effectiveness of \name{}}
\label{sec:performance of anofusion}
Table~\ref{table:baselines anomaly detection result} lists the average precision, recall, and best \fscore{} of \name{} and seven baseline approaches described above on the two datasets.
\name{} outperforms all baseline approaches on both datasets, with best \fscore{} of 0.857 and 0.922, respectively.

\point{Multimodal failure detection}
SCWarn performs better than single-modal failure detection methods by simultaneously processing metrics and logs on D2.
However, it ignores tracing data, which is crucial for detecting instance failures in microservice systems on D1.
The correlation among each modality is ignored by JLT, yielding sub-optimal performance on both datasets.

\point{\name{}}
Our approach is effective to detect instance failures, with the average best \fscore{} significantly higher than existing methods.
Compared with SCWarn which combines two modalities, the average best \fscore{} of \name{} outperforms it by 41.00\% and 18.80\% on both datasets, respectively.
Compared with JLT, using a heterogeneous graph stream significantly improves the effectiveness of instance failure detection. 
\name{} outperforms JLT  by 23.90\% and 44.10\% on both datasets, respectively.

\point{Robustness comparison}
Firstly, we use two datasets from different microservice systems, and the experiments show that \name{} achieves superior detection results on both datasets, outperforming all baselines.
Secondly, each dataset contains many kinds of instances. 
We analyze the failure detection results of each instance for each dataset.
The \fscore{}s of \name{} for all instances on D1 range from 0.784 to 0.977, and from 0.805 to 0.986 on D2.
We can see that \name{} performs well in both D1 and D2.
Therefore, we believe these are testaments to the robustness of \name{}.

\point{Efficiency comparison}
We simulate the online detection environment and analyze the complexity of \name{} and other baselines by counting the detection time required for each sliding window.
\name{} takes a window containing the data points as input and then calculates a failure score through the trained model.
The average running time of \name{}'s online failure detection is $1.932 \times 10^{-2}$s.
The prediction time of other baselines is $9.810 \times 10^{-3}$s for JumpStarter, $8.975 \times 10^{-5}$s for USAD, $1.707 \times 10^{-3}$s for LogAnomaly, $1.165 \times 10^{-4}$s for Deeplog, $1.102 \times 10^{-2}$s for TraceAnomaly, $3.331 \times 10^{-3}$s for SCWarn, and $9.958 \times 10^{-3}$s for JLT.
Since operators perform failure detection every minute, every approach can satisfy this requirement.
Furthermore, \name{} achieves satisfactory results by leveraging the three modalities, which is superior considering both effectiveness and performance.

\subsection{RQ2:  Contributions of Components}
\label{sec:evaluation_ablation_study}

To demonstrate the contribution and importance of each component of \name{}, we create three variants of \name{} and conduct a series of experiments to compare their performance. These variants are:
\begin{enumerate*}[1)]
    \item \textbf{\name{} w/o GTN.} To study the significance of GTN in modeling the heterogeneity and correlation of multimodal data, we remove GTN from \name{}.
    \item \textbf{\name{} using GCN.} To show the importance of assigning different weights to neighbor nodes in the graph (attention mechanism), we use GCN instead of GAT in \name{}.
    \item \textbf{\name{} w/o GAT.} To demonstrate how the graph attention mechanism improves \name{}'s performance, we remove the GAT from \name{}.
\end{enumerate*}

\begin{table}[t]
    \centering
    \caption{
        The average precision, recall, and \fscore{} of \name{} and different model variants
    }
    \label{table:Abortion}
    \begin{tabular}{c|c|c|c|c}
        \toprule
        & \textbf{Approach} & \textbf{Precision} & \textbf{Recall} & \textbf{\fscore{}}\\
        \midrule
        \multirow{4}{*}{D1} & w/o GTN & 0.608 & 0.891 &  0.723 \\
        & use GCN & 0.769 & 0.847 & 0.802 \\
        & w/o GAT & 0.602 & 0.643 & 0.615\\
        & \name{} & \textbf{0.795} & \textbf{0.945} & \textbf{0.857} \\
        \midrule
        \multirow{4}{*}{D2} & w/o GTN & 0.742 & 0.917 &  0.859 \\
        & use GCN & 0.819 & 0.863 & 0.823 \\
        & w/o GAT & 0.698 & 0.735 & 0.700 \\
        & \name{} & \textbf{0.863} & \textbf{0.991} & \textbf{0.922} \\
        \bottomrule
    \end{tabular}
\vspace{-0.2cm}
\end{table}


Table~\ref{table:Abortion} lists the average precision, recall, and best \fscore{} of the three variants discussed above on two datasets.
When GTN is removed, both precision and recall are degraded.
The decrease in precision is especially obvious.
It shows that GTN can capture heterogeneity and correlation among multimodal data in heterogeneous graphs, thereby reducing false positives by comprehensively synthesizing the information of the three modalities.
Both precision and recall decrease when GCN is substituted with GAT, demonstrating that GAT is more efficient than GCN for dynamically changing time series.
Each node in the heterogeneous graph stream behaves differently.
Treating all nodes equally like GCN introduces noise to the graph stream that can interfere with the model training process.
Furthermore, when GAT is no longer present in \name{}, the precision and recall drop dramatically, which indicates that GAT can focus on the most relevant information in the graph.


\subsection{RQ3: Hyperparameters Sensitivity}
\label{subsec:eva hyper-parameters}

\begin{figure}[t]
    \centering
    \includegraphics[width=\linewidth]{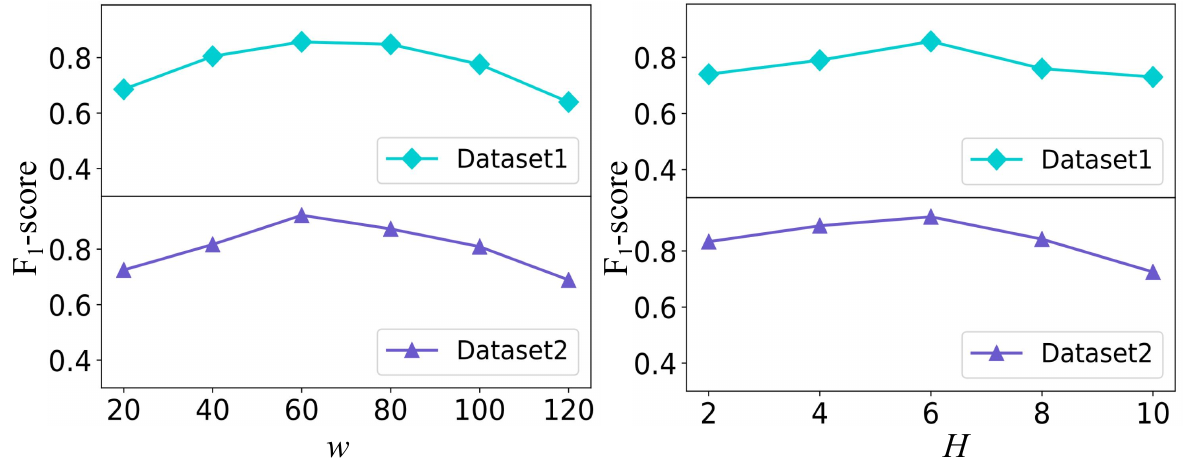}
    \caption{
        \fscore{} of \name{} under different parameters
    }
    \label{fig:para}
\end{figure}

We mainly discuss the effect of two hyperparameters in multimodal data serialization and graph stream construction on \name{}'s performance.
Figure~\ref{fig:para} shows how the average best \fscore{} of \name{} changes with different values of hyperparameters.
Specifically, we increase the size of the sliding window in data serialization, $\theta$, from 10 to 120.
From the experiment results, we can find that if $\theta$ is too large, it will contain too many seasonal variations and will struggle to reconstruct the current state; if $\theta$ is too small, the model will be unable to comprehensively learn the information from historical data, degrading \name{}'s performance.
$\theta$ between 40 and 90 can lead to relatively good performance.
Thus, we set $\theta=60$.

We change the number of attention heads in GAT, $H$, from 2 to 10.
An $H$ of 6 yields the best performance.
If $H$ is too small, the performance of \name{} will slightly degrade due to the decrease in model size; if $H$ is too large, more redundant information may be generated, interfering with the training of \name{}~\cite{DBLP:conf/acl/VoitaTMST19}.

\section{Discussion}
\subsection{Lessons Learned}

\noindent \textbf{Collecting multimodal monitoring data in real-time.} 
We utilize multimodal data to detect failures in instances.
Ensuring the real time data quality of different modalities is essential for the deep learning models. 
From our real-world experience in Microsoft, we suggest leveraging the open-source monitoring systems or Azure Monitor\footnote{\url{https://azure.microsoft.com/en-us/products/monitor}} to build the data pipeline. 
For example, Prometheus\footnote{\url{https://prometheus.io}} can be used to collect metrics.
ELK (Elasticsearch, Logstash, and Kibana) Stack\footnote{\url{https://www.elastic.co/what-is/elk-stack}} are used to collect logs.
Skywalking\footnote{\url{https://skywalking.apache.org}} can be used to collect traces.
Additionally, 16 days of data are utilized for training.
When the model training is completed, \name{} digests 10 minutes of data to perform real-time detection in the online detection stage, which is efficient and effective in practice.

\noindent \textbf{Selection of evaluation metrics.}
In the online detection stage, \name{} adopts the EVT algorithm~\cite{siffer2017anomaly} to obtain the best \fscore{}.
In practice, however, operators may have varying preferences for precision and recall depending on the business type.
For example, operators generally seek a high recall for the core services that provide online shopping services. 
They do not want to miss any potential failures that could negatively impact users' experience.
Precision is often preferable in data analytic services.
Operators want to detect failures accurately rather than receive a large number of false alarms.
Therefore, concentrating solely on \fscore{} is not appropriate for all instances.
In the future, we plan to provide an interface that allows operators to apply additional weights, valuing one of precision or recall more than the other.

\subsection{Threat to Validity}
\label{sec:threat to validity}
\point{Failure labeling}
In our experiments, we use two datasets, one is public and another is collected from a real-world commercial bank.
The ground truth labels are based on failure injection (D1) and manually checking failure reports by system operators (D2). 
Manually labeling anomalies may contain few noises. 
We believe that the labeling noises are minimal due to the extensive experience of operators.
Furthermore, to reduce the impact of labeling noises, we adopt widely used evaluation metric to detect continuous failure segments instead of point-wise anomalies \cite{ma2018robust}.

\point{Granularity effect}
The granularity of the monitoring data in our experiments is one minute, but this has no effect on the algorithm's effectiveness. 
We believe the algorithm can still work with finer or coarser-grained data without additional effort. 
The datasets in our experiments are still limited.
We will experiment \name{} with a larger scale of system in the future.

\point{Data modalities}
Our work involves the utilization of three modalities of monitoring data.
We believe that in a real-world scenario, as long as the modalities of the monitoring data are no less than two, the algorithm will function normally. 
Furthermore, if a failure manifests itself in only one type of monitoring data, \name{} will consider not only the correlation among historical multimodal data, but also the proportion of anomalous in all monitoring data to determine whether the instance fails.

\section{Conclusion}

Failure detection in the microservice systems is of great importance for service reliability. 
In this work, we propose \name{}, one of the first studies using multimodal data, \textit{i.e.}, metrics, logs, and traces, to detect failures of instances in microservice systems robustly.
Specifically, we first serialize the data of the three modalities and construct a heterogeneous graph structure.
Then, GTN is utilized to update the heterogeneous graph, with GAT being used to capture significant features.
Finally, we use GRU to predict the data pattern at the next moment.
The deviation between the predicted and observed values is used as the failure scores.
We apply \name{} on two microservice systems, which proves that it significantly improves the \fscore{} for failure detection.
We believe that the solution of applying multimodal data for failure detection will benefit more areas beyond microservice systems.

\begin{acks}
    The work was supported by the National Natural Science Foundation of China (Grant No.~62272249 and 62072264), the
Natural Science Foundation of Tianjin (Grant No. 21JCQNJC00180), and the Beijing National Research Center for Information Science and Technology (BNRist).
\end{acks}

\clearpage
\bibliographystyle{ACM-Reference-Format}
\balance
\bibliography{main}

\end{document}